# Enhanced spontaneous emission from nanodiamond colour centres on opal photonic crystal

FA Inam[1,2,3], T Gaebel[1,2,3], C Bradac[1,2,3], L Stewart[2,4], MJ Withford[1,2,4], JM Dawes[2,4], JR Rabeau*[1,2,3] and MJ Steel[#1,2,4]

1. Centre for Quantum Science and Technology, Department of Physics and Astronomy, Macquarie University, Sydney, New South Wales 2109, Australia

2. MQ Photonics Research Centre, Department of Physics and Astronomy, Macquarie University, Sydney, New South Wales 2109, Australia

3. ARC Centre of Excellence for Engineered Quantum Systems, (EQUS)

4. ARC Centre of Excellence for Ultrahigh-bandwidth Devices for Optical Systems, (CUDOS)

*james.rabeau@mq.edu.au

[#]michael.steel@mq.edu.au

**Abstract.** Colour centres in diamond are promising candidates as a platform for quantum technologies and biomedical imaging based on spins and/or photons. Controlling the emission properties of colour centres in diamond is a key requirement for developing efficient single photon sources with high collection efficiency. A number of groups have produced enhancement in the emission rate over narrow wavelength ranges by coupling single emitters in nanodiamond crystals to resonant electromagnetic structures. Here we characterise in detail the spontaneous emission rates of nitrogen-vacancy centres positioned in various locations on a structured substrate. We show an average factor of 1.5 enhancement of the total emission rate when nanodiamonds are on an opal photonic crystal surface, and observe changes in the lifetime distribution. We present a model to explain these observations and associate the lifetime properties with dipole orientation and polarization effects.

## 1. Introduction

In order to meet some of the demanding requirements for optical quantum technologies, bright sources of single photons [1-5] are required. Colour centres in diamond nanocrystals are among the leading candidates to fill this need, due to a host of attractive properties including room-temperature photostability and convenient spin polarization by optical pumping. However, existing realisations are far from ideal—intrinsic spontaneous emission lifetimes in the range 1-30 ns combined with the problem of maximising the collection efficiency of emitted photons sets a maximum limit to emission rates of a few MHz. In addition, at least for the most widely studied nitrogen vacancy (NV) centre, the intrinsic emission spectrum is very broad. Consequently, several groups are presently developing techniques to enhance both the emission rate and photon collection efficiency, as well as to manipulate the spectral shape.

A variety of approaches have been explored but all are based on the idea of changing the local electromagnetic environment seen by the emitted photons. One class of methods involves coupling the radiating dipole to a single strongly-localised mode of a high $Q$ resonator such as a microsphere [6, 7], microdisk [8] or photonic crystal cavity [9-12], enhancing the emission rate through the Purcell effect [13, 14]. However as mentioned, a key challenge of diamond colour centres, as compared to other systems such as quantum dots, is the linewidth of the intrinsic emission spectrum. For the NV centre, the overall linewidth is of order 200 nm with the zero phonon line (ZPL) at 637 nm having a width of several nm at room temperature [15, 16]. By contrast, the cavity linewidth of a typical photonic crystal cavity with a $Q$ of $10^4$ is over an order of magnitude smaller than the width of the ZPL. Therefore while the rate of emission within the cavity line can increase markedly, this represents a small fraction of the total emitted power. So the primary effect of coupling the centre to a cavity mode is a reshaping of the emission spectrum, with little increase in the overall rate of emitted photons. Additional increase in the cavity $Q$ increases the local emission rate further, but also reduces the frequency range over which enhancement occurs by the same factor. Some improvements are possible: under cryogenic conditions, the width of the zero phonon line does narrow [17] (though it remains significantly larger than the cavity linewidth) and therefore a slightly stronger effect is expected. Alternatively, one could use other centres such as the nickel-related [18-22] group which have narrower intrinsic spectra (though again still broader than the cavity linewidth), however the fabrication and spectroscopy of these centres are not as mature as for NV centres.

Consequently, there is merit in pursuing other complementary schemes that enhance the emission rate over a broad spectral band at the expense of very strong enhancement for a single wavelength, and to investigate influences on the *distribution* of emission rates. One recent example exploited the very large local electric fields associated with localized surface plasmons to achieve enhancement in overall emission rate of an NV centre by a factor as large as 14 [23, 24]. We recently showed less pronounced but broad band *suppression* in the emission rate by embedding nanodiamonds in regular 3-D photonic crystals [9].

For the most part, these considerations apply to emission by a single centre with a given orientation for its electric dipole moment, and a given nanodiamond size. If an ensemble of measurements is considered, the situation becomes more complicated as there is variation in the emission rate from centre to centre associated with nanodiamond size and shape [25], as well as the orientation of the centre dipole with respect to the local electromagnetic environment (the refractive index distribution). The relative importance of these different contributions also depends on the type of structure with which the nanodiamond colour centre is interacting, for example, a flat interface or structured surface. A recent work reported on the lifetime distributions of NV centres in disordered scattering media [26]. The lifetime distribution of NV centres in nanodiamond crystals was measured on a coverslip surrounded by air and subsequently by a droplet of water, and a small reduction in the mean lifetime was observed. Finally the nanodiamond crystals were surrounded with high refractive index dielectric spheres of sizes around 250 nm and a lengthening in the fluorescence lifetime was observed which was attributed to a reduction in the local density of states (LDOS).

In this work, we theoretically and experimentally study emission of single NV centres placed on flat dielectric substrates, and on a structured opal photonic crystal surface. We consider the statistics of lifetimes associated with size, shape and orientation and measure a 1.5-fold reduction in the overall lifetime of single nitrogen vacancy centres on the surface of the opal. The enhancement is interpreted in terms of an averaging of the polarisation dependence of the emission rate associated with the structured surface. The remainder of the paper is structured as follows. In section 2 we review the theory of dipole emission rates in structured dielectric environments based on the local density of states, and discuss issues peculiar to NV nanodiamonds. The experimental details starting from sample preparation to the characterisation and lifetime measurement of single NV centres are described in section 3. The emission rate calculations are presented and compared to experimental results in section 4. Section 5 discusses the sensitivity of the calculations with respect to non-idealities in the nanodiamonds and we make conclusions in section 6.

## 2. Theoretical background of emission rates in structured dielectrics

We begin by recalling the treatment of emission in structured environments and discussing issues related to NV nanodiamonds. In the weak coupling regime that applies in this work, the spontaneous emission rate $\Gamma$ of a dipole at position $\mathbf{r}$ is determined by Fermi's golden rule applied to electric dipole transitions [27]:

$$\Gamma = \frac{1}{\tau} = \frac{\pi\omega}{\hbar}\mu^2 \sum_{\mathbf{k},\sigma}\left|\hat{\mathbf{d}}\cdot\mathbf{E}_{\mathbf{k},\sigma}(\mathbf{r})\right|^2 \delta(\omega-\omega_{\mathbf{k},\sigma}), \qquad (1)$$

where $\tau$ is the spontaneous emission lifetime, $\omega$ is the transition frequency, $\boldsymbol{\mu} = \mu\hat{\mathbf{d}}$ is the relevant dipole moment matrix element and the sum is taken over all possible emitted photon modes $(\mathbf{k},\sigma)$ labelled by wavevector and polarisation. The modes have local electric field profiles $\mathbf{E}_{\mathbf{k},\sigma}(\mathbf{r})$. In particular, the rate is influenced by the dielectric environment through the electromagnetic local density of states (LDOS) [28-30]

$$\rho(\omega,\mathbf{r}) = \sum_{\mathbf{k},\sigma}\left|\hat{\mathbf{d}}\cdot\mathbf{E}_{\mathbf{k},\sigma}(\mathbf{r})\right|^2 \delta(\omega-\omega_{\mathbf{k},\sigma}). \qquad (2)$$

The LDOS accounts for both the local field intensity of the available modes and the relative orientation of the dipole with respect to the field of each mode. The LDOS itself is a purely classical quantity that can be found from the electromagnetic Green tensor. This exact picture encompasses a range of systems. At one extreme is emission enhancement by coupling to a single high $Q$ cavity mode of large field amplitude and small mode volume. At the other extreme of a dipole embedded in a homogeneous medium, the dipole couples to a continuum of modes. The emission rate is then proportional to the refractive index $n$ with an $n^3$ increase in the density of modes outweighing a reduction in the electric field intensity of $n^2$:

$$\Gamma(n) = n\Gamma_0. \qquad (3)$$

The parameter $\Gamma_0$ is the emission rate of the dipole in free space. (Here we neglect an atomic-scale local field factor which for our purposes can be regarded as a constant since the NV centre is locally surrounded by the same diamond lattice whether we consider bulk or nano-diamond [31, 32].) For both these extremes, the dipole orientation plays a trivial role, with a cosine factor in the case of a single cavity mode, and no dependence at all for a homogeneous medium.

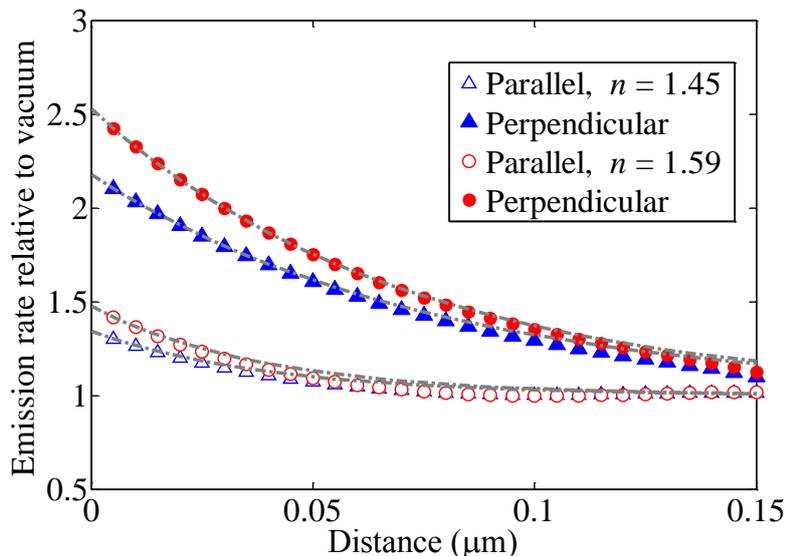

Figure 1: Relative emission rate for radiating dipoles near an interface between two dielectrics, corresponding to parallel and perpendicular dipole orientations relative to the interface, for an emission wavelength of 680 nm. The dashed lines represent the semi-analytical theory in [33]. The markers are exact FDTD results. The small discrepancy at large distances is due to an approximation in the analytic theory. Curves are labelled according to the relative index constant $n = n_2/n_1$.

In contrast, the emission rate of dipoles near nonuniform but non-resonant structures is affected by both the distribution of photonic modes and the dipole orientation. This set of structures is exactly the class that may help to induce broadband enhancement of dipole emission. The simplest example is a dipole on or near a flat dielectric surface, a system which has been studied in detail over many years [33, 34]. In this case, the emission rate is determined by interference between directly emitted waves and those reflected from the surface. The relative emission rate as a function of distance from the interface is shown in Fig. 1 for parallel and perpendicular dipole orientations and for two different substrate refractive indices. There is a strong dependence on the dipole orientation through the polarisation dependence of the Fresnel reflection coefficients.

## 2.1 Local dielectric environment

The emission of NV centres in nanodiamonds is commonly characterised for crystals deposited on a glass substrate [33, 35], and in the first part of our experiments below we also make such a measurement. The nanodiamond on a substrate might seem to correspond to the geometry just described of a dipole near a surface, but as we discuss below the fact that the dipole is trapped inside a small diamond crystal has a crucial effect. In fact for this problem, several reports in the diamond literature [33, 35] have invoked an even simpler scalar picture, arguing that the dipole effectively emits into two half spaces each of which promotes emission at the rate $n/(n_d \tau_b)$, according to Eq. (3). Here $n_d \approx 2.4$ is the refractive index of diamond and $\tau_b = 11.6$ ns is the accepted spontaneous emission lifetime for an NV centre in bulk diamond. Therefore the resulting lifetime should be approximately given by the expression [35]

$$\tau_{sub} = \frac{1}{\frac{1-f_s}{n_d \tau_b} + \frac{n_s f_s}{n_d \tau_b}}, \tag{4}$$

with the volume fraction $f_s = 1/2$, $n_s$ being the substrate index. This expression does agree quite well with the measured mean lifetime of around 23 ns [33, 35]. Figure 2 provides an extension to this treatment allowing the dipole to be embedded in non-planar two-component structures. The figure shows the expected lifetime for a single NV centre as a function of the volume fraction $f_s$ of high index material in its surroundings for several relevant substances. Quite simply, as the dipole emitter becomes more enveloped in material, its lifetime decreases. While this ignores interference effects and dipole orientation, it provides some context for our subsequent more rigorous results.

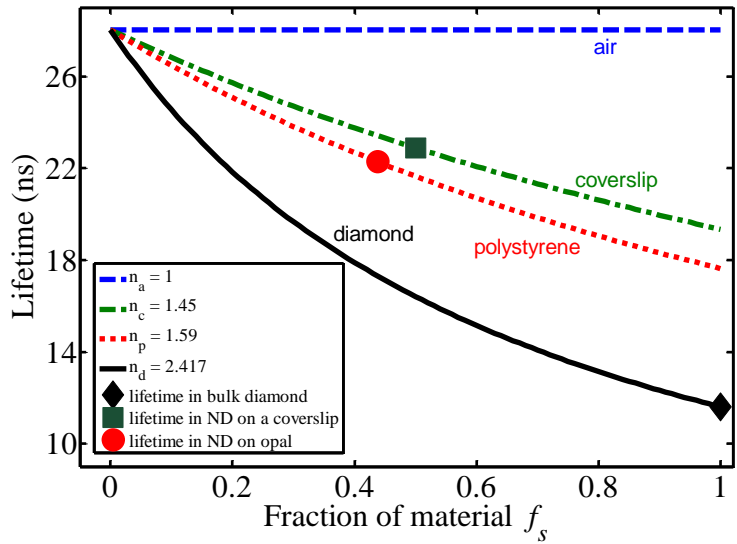

Figure 2: Lifetime for a single NV centre as a function of the volume fraction of high index material in its surroundings for several relevant materials [35]. The circle, square and diamond symbols show the lifetime of the NV centres on coverslip and opal substrates inside a nanodiamond (ND) crystal and in bulk diamond, respectively.

Although the argument leading to Eq. (4) is reasonable for an isolated dipole such as a dye molecule close to a surface, it is not appropriate for a dipole trapped inside a high-index sub-wavelength particle, such as an NV centre in a nanodiamond with diameter $a < 100$ nm. It has long been known that for such nanoparticle-embedded emitters, where the emission wavelength $\lambda$ is larger than the particle, the emission rate $\Gamma_p$ is expected to be suppressed compared to the rate $\Gamma_b$ in a macroscopic piece of the same material [25, 26, 31, 36]. For a spherical particle in the limit $\lambda \gg a$, the suppression ratio tends to

$$\frac{\Gamma_p}{\Gamma_b} = \frac{1}{\varepsilon}\left(\frac{3}{\varepsilon+2}\right)^2 \approx 0.062, \qquad (5)$$

where $\varepsilon = n_d^2$. The treatment of Chew [25] shows that the suppression is essentially a local field effect at the scale of the particle with the electric field inside the nanodiamond reduced by the Lorentz-Lorenz factor $3/(\varepsilon+2)$. (This is distinct from the atomic-scale local field effect the dipole experiences due to the diamond crystal lattice structure [37]. As mentioned earlier, since in both bulk diamond and nanodiamond the NV centre is embedded in a diamond lattice, we expect the atomic local field factor to be unchanged.) Emission suppression in small particles has been clearly demonstrated for dye molecules in polystyrene spheres by Schniepp and Sandoghdar [36]. At least for a spherical particle [25], the emission rate depends very weakly on the precise position and orientation of the dipole inside the particle. For non-spherical particles, some dependence on orientation leading to a broadening of the linewidth distribution is expected [32, 38]. We show in section 5.2 that this is a minor effect in our systems.

It follows from this discussion that a correct theory of emission from a colour centre within a nanodiamond on any kind of substrate, whether planar or otherwise, must explicitly incorporate the influence of the diamond material in the nanoparticle as well as the surrounding environment. Moreover, the dependence on orientation and position from the interface as suggested by Figure 1 indicates that a much more sophisticated treatment than Eq. (4) is required to fully understand the emission behaviour. Such a treatment seems to be uncommon in the literature and complicates the calculation as we discuss in the next section. Ruijgrok *et al.* [26] report having performed finite-difference time-domain (FDTD) simulations but do not provide details.

## 2.2 Structured surfaces

In the second part of our experimental study we consider emission from diamond nanocrystals placed on the surface of an opal photonic crystal composed of polystyrene spheres with refractive index 1.59 (see Fig. 3). An opal photonic crystal consists of a regular 3-dimensional arrangement of microspheres in a face centered cubic structure [39, 40]. We have previously demonstrated emission suppression for nanodiamonds embedded *within* an opal photonic crystal with a stop band tuned to the emission band [9]. In the present work, we concentrate on the change in emission rates due to the structured surface of the opal rather than through photonic band gap effects.

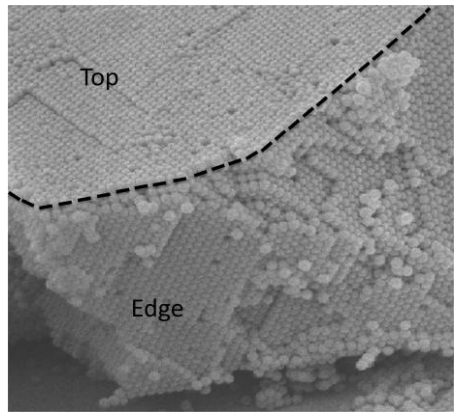

Figure 3: Scanning electron microscope (SEM) image of the opal photonic crystal.

Specifically, the nanodiamonds can be positioned on the apex of a polystyrene bead or in an interstitial position between three microspheres thus presenting a range of interaction conditions depending on the surrounding environment. The fraction of nearby space occupied by the high index material varies from site to site, but just as importantly, so does the relative orientation of adjacent surfaces. In particular, for a nanocrystal positioned in an interstitial site, the distinction between parallel and perpendicular orientation to nearby boundaries is much less pronounced than for the flat surface (see Fig. 4) and we might expect some averaging of these two extremes with a net increase in emission rate. We show below that the measured and predicted increase considerably exceeds the change that would be predicted based only on the variation in material volume fraction illustrated in Figure 2.

## 3. Experimental details
We now describe our experiments including preparation of nanodiamond doped substrate samples and lifetime measurements.

### 3.1 Sample Preparation
Coverslips of thickness (150 ± 20) μm were etched in piranha acid solution ($H_2SO_4$:$H_2O_2$, 3:1) and dried in a dry nitrogen flow. Opal films were grown using the controlled evaporation method [41] which we briefly describe here. A coverslip was placed into a scintillation vial containing a 10 mL solution of polystyrene microspheres (0.320 ± 0.016) μm diameter, (Bangs Laboratories) diluted by ultra-pure water. The refractive index of the microspheres was specified to be $n_p$=1.59. Dilutions of 0.1% weight of polystyrene were used. The vial was then placed in a temperature controlled oven, set to 40°C [9]. Coverslip and opal samples were prepared by pipetting a 100 μL ethanol suspension of diamond nanocrystals (containing 7 mg diamond nanocrystals per millilitre) over each surface and allowing it to dry. These diamonds were previously analysed by atomic force microscopy [42, 43] and found to have a mean size of 54 nm and a distribution of 30-80 nm. Nanodiamonds deposited on the opal were imaged using scanning electron microscopy in order to assess the typical distribution of nanodiamonds over the opal surface. Of the 365 nanodiamonds analysed, $f_T = 10\%$ of the particles tended to be located on the top of a bead, $f_I = 65\%$ were in the interstitial position and the remaining $f_R = 25\%$ were distributed in various other locations in between the top and the interstitial position (see Figure 4b).

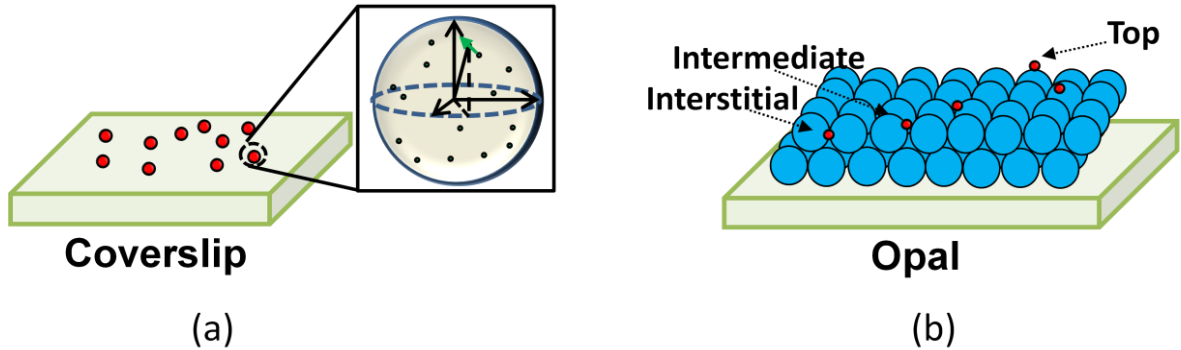

Figure 4: Schematic representation of nanodiamond positions on (a) the coverslip and (b) the opal layer which were experimentally and computationally studied. The inset shows some of the possible random positions (dots) of the dipole inside the diamond sphere as well as a possible random orientation of an individual dipole (arrow).

### 3.2 Lifetime measurements
A 100mW continuous wave laser (Coherent Compass ~300 μW incident on the sample) operating at 532 nm was used to optically excite luminescence from the NV centres. The laser was focused through the back surface of the coverslip and onto the sample using a 100× infinity-corrected oil immersion objective with a numerical aperture of 1.3 (Olympus) and luminescence was collected confocally through a pinhole. The diamond was excited at normal incidence to the $\langle 111 \rangle$ plane of the opal, which naturally grows parallel to the substrate. A spectrometer (Acton) with a cooled CCD (Princeton Instruments) was used to characterize the luminescence and a Hanbury Brown and Twiss (HBT) interferometer with single-photon-sensitive avalanche photodiodes (Perkin Elmer SPCM-AQR-14) was used to measure the photon statistics. Photon counting and correlation was carried out using a time-correlated single-photon-counting (TCSPC) module (PicoHarp 300, PicoQuant GmbH). To eliminate the complication of multiple NV lifetimes in a single crystal, we identified diamonds with single NV centres by measuring the second order correlation function $g^{(2)}$ using the HBT setup. A dip in the $g^{(2)}$ curve below 50% of the maximum value indicates a single centre as seen in Figure 5 (a), (b), (d) and (e). Contrast below 100% is due to the background signal (reflection of the laser or emission from a nearby centre) [35]. Based on our previous characterisations [42], less than 1% of nanodiamonds used in these experiments are expected to contain single NV centres. In this case, we identified a total of $N_C = 37$ single centres on the coverslip and $N_O = 27$ single centres on the opal. Unfortunately, the opal surface does not appear in our confocal images, so we could not identify the specific environment (top, interstitial, other) of each individual centre. In our calculations below, we assume they are distributed with similar statistics to the full set of 365 nanodiamonds observed with the SEM, as described in section 3.1.

For each single NV centre, $g^{(2)}$ correlation traces were collected over a range of laser powers. The $g^{(2)}$ correlation traces (see Fig. 5 a, b, d, e) were fitted to a three-level model for the optical transition to obtain the decay times for the transitions corresponding to different excitation powers. The lifetime of the NV centre was then inferred by extrapolating the value of the decay time to vanishing excitation power [16, 35]. Note that the powers indicated in Figure 5 are input powers and the much larger values for the opal are due to the need to excite the NV centre through the opal substrate. The collected photon counts and thus the excitation power at the site of the nanodiamond were of the same order for both coverslip and opal [16, 35].

The measured lifetime distributions are shown in Figure 6. The mean lifetime for the diamonds on a coverslip was $(25.4 \pm 1.7)$ ns and for the diamonds on opal was $(17.5 \pm 1.6)$ ns (uncertainties are the standard error of the mean). As a representation of the distribution width, in Figure 6 we also provide the range containing the middle 50% of measured lifetimes. The change from the coverslip to the opal produced a lifetime reduction of $\Delta \tau = (7.9 \pm 3.3)$ ns. To quantify the difference in the distributions we performed a Kolmogorov-Smirnov test [44] and found that the probability that the two sample distributions come from the same underlying distribution was less than 1%.

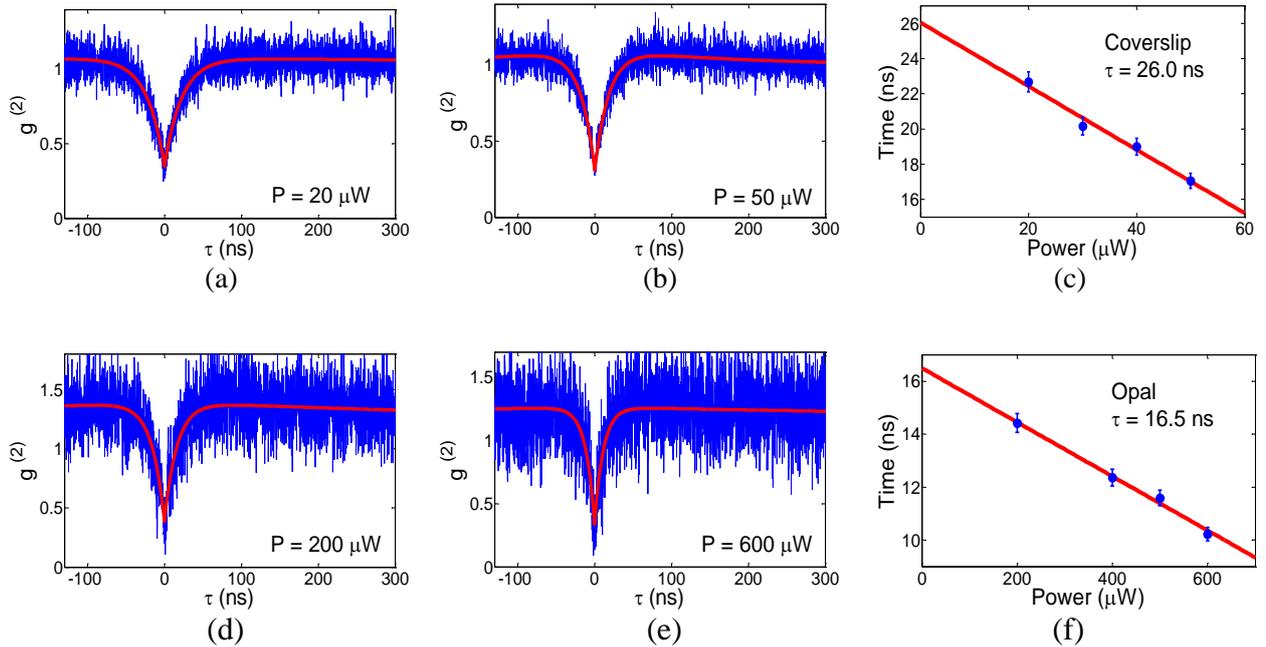

Figure 5a), b), c), and d): Typical background corrected second order correlation measurements (blue) and fit (red) to a 3-level system [16] to obtain decay times for different excitation powers. e) and f): Lifetimes as a function of input power. The final lifetime value was obtained by extrapolating the decay time to vanishing excitation power [16]. The error bars in the figure correspond to a 5% fluctuation in the excitation power.

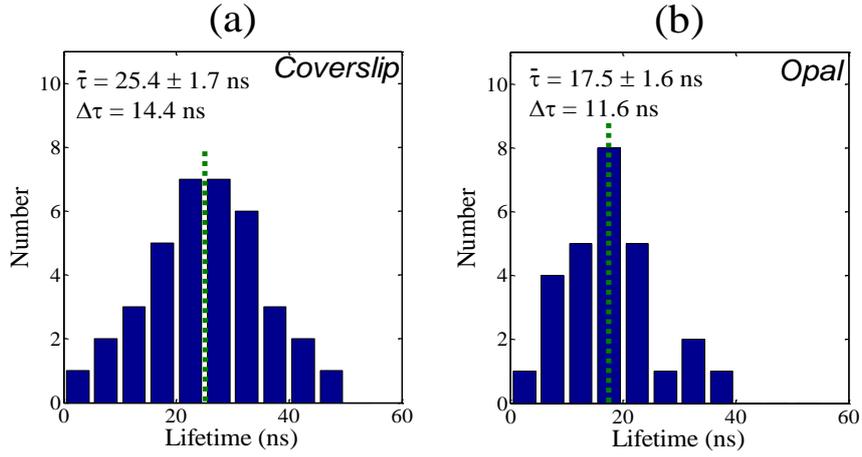

Figure 6: Experimental distribution of the NV centre lifetime on a coverslip a) and on an opal b). Dotted lines denote the mean value. Quoted values are mean lifetime $\bar{\tau}$ and the range $\Delta\tau$ that contains 50% of centres.

## 4. Emission rate calculations

We now present a model for the emission rate distributions for the coverslip and opal geometries. Although spontaneous emission is an explicitly quantum process, the emission rate relative to a reference system can be found by classical electromagnetic calculations since the dependence on environment is expressed through the classical LDOS [see Eq. (2)]. For each geometry discussed above, NV emission rates relative to the rate in bulk diamond were calculated using 3D finite-difference time-domain (FDTD) simulation as follows.

### 4.1 Method and geometry

The dipole was considered to be enclosed in a spherical nanodiamond crystal with a diameter $D$=54 nm. We discuss the sensitivity of our results to shape and size of the nanodiamond in section 5. The nanodiamond was placed in contact with either a glass substrate or a part of the opal surface. The index of the glass coverslip was 1.45. The radius of the opal spheres was $a = 160$ nm and the refractive index was 1.59. For a given dipole position and orientation within the nanosphere, the electromagnetic field was excited by a sinusoidal point current source driven at a frequency $\nu = c/\lambda$ for $\lambda = 680$ nm which is near the peak of the NV emission. Once the field configuration reached steady-state, which takes only a few optical periods as the reflected fields are weak, the power $P$ radiated by the dipole was calculated by two methods—finding the work done by the dipole $W = \langle \mathbf{J}(t) \cdot \mathbf{E}(t) \rangle$ averaged over one optical cycle, and integrating the time averaged Poynting flux $\mathbf{S} = \langle \mathbf{E}(t) \times \mathbf{H}(t) \rangle$ over the surface of a rectangular box enclosing the dipole. The calculation was then repeated with the same dipole orientation in a uniform medium of index $n_d = 2.4$ to obtain the emitted power $P_b$ corresponding to NV emission in bulk diamond. The spontaneous emission rate enhancement was then found as $R = P/P_b$.

The calculations were performed with a spatial grid of $\Delta x = 5$ nm, for which the two methods of calculating radiated power were in close agreement. The total domain dimensions were 300 nm on a side for the coverslip and 700 nm for the opal and the Poynting flux integration box had dimensions 270 nm and 630 nm respectively. We checked convergence by repeating some simulations on grids of $\Delta x = 2.5$ nm and $\Delta x = 1.25$ nm, and found the time-averaged Poynting flux was well converged. The $\mathbf{J}(t) \cdot \mathbf{E}(t)$ method became unreliable at small grid sizes, as it becomes more sensitive to the pole in the Green function at the source point.

To confirm the accuracy of our calculations, we checked that our simulation method reproduced the analytic results of Chew [25] for the spontaneous emission rate of a dipole inside a sphere alone. Figure 7 shows a comparison between the analytic results and our FDTD calculations for a dipole displaced by a factor of $0.6a$ from the centre with a grid size of $\Delta x = 5$ nm. Our results agree with the analytical calculations out to $2a/\lambda = 0.5$, ie. for wavelengths much shorter than those used in our main calculations, which correspond to the range indicated by the solid green bar. Similarly we compared our FDTD method with the analytical result for the case of a dipole near a planar interface as shown in Fig. 1. The FDTD results (dashed lines) agree well with the semi-analytic theory (dots).

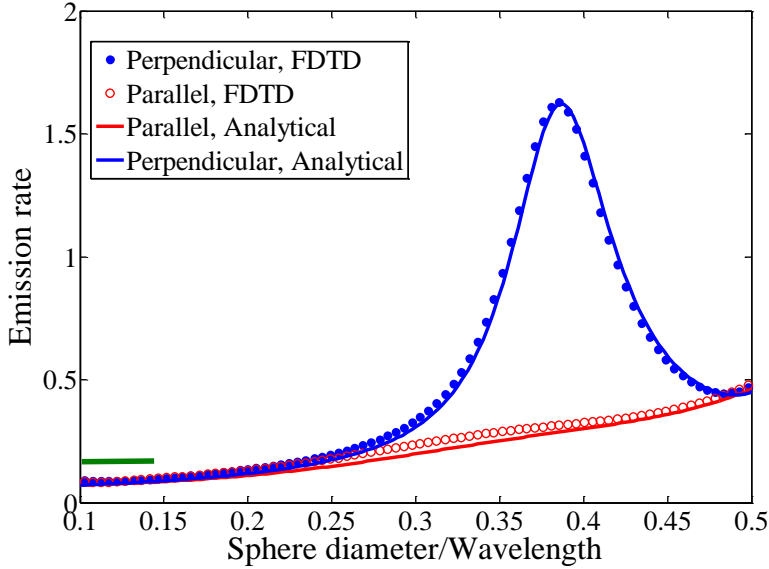

Figure 7: Comparison of emission rates for a dipole inside the diamond sphere (both perpendicular and parallel orientations) calculated using FDTD and the analytical approach [25]. The emission rates are relative to the rate in bulk diamond. The solid green line corresponds to the size distribution of the nanodiamond crystals used in our experiments.

The simulations were performed for the diamond resting on the coverslip, and for three different configurations on the opal: the diamond in the centre of the interstitial site; the diamond resting 20 nm above the interstitial position; and the diamond sitting on top of one sphere in the opal microstructure (see Figure 4). The latter three positions were chosen on the basis of the SEM measurements discussed in section 3. The interstitial position was taken as the centre position between the three surrounding spheres with the size of diamond considered to just span the gap between the adjacent beads. With variation in the diamond dimension, this position varies slightly, though the variation was found to have less than a 5% effect on the overall emission rate for a 10 nm fluctuation in diamond position. Similarly, the site 20 nm above the interstitial is considered to be representative of the distribution of intermediate locations. The low number of experimental samples and the resolution of the SEM measurements makes a more precise comparison impossible.

### 4.2 Statistical averaging

For each of these four configurations, we need to average over the position and orientation of the dipole within each nanodiamond, since in our experiments we have no control over either variable. We assume that the dipole is uniformly distributed within the spherical nanodiamond (excluding a 5 nm surface shell) [45], and that the dipole orientation has no preferred direction, giving a total of five degrees of freedom. In calculations of emission rates for ideal systems, it is often possible to decompose the dependence on dipole orientation into emission due to dipole oscillation in two or three uncoupled perpendicular directions. This is true for both a dipole in a sphere surrounded by uniform medium [25] and for a dipole above a plane interface [33]. The overall emission rate is an angle weighted sum of the component rates. In general however, due to low symmetry there can be coupling of the radiated electric field from one dipole orientation to the dipole motion in another. This is the case for our situation of a dipole inside a sphere which itself is adjacent to a plane. The complete dependence on angle can then only be found by a sufficiently dense sampling of all directions on the unit sphere. Therefore to completely characterise the expected distribution of emission rates we would need to perform a double average over all possible dipole positions and all possible dipole orientations at each position. Not only would this be a very large numerical undertaking but it seems excessive since our experiments involve ensembles of only around 30 observations and the experimental statistics are consequently quite noisy. Instead, to obtain a manageable number of simulations, we have constructed distributions of the lifetime by performing 1000 simulations with uniformly distributed random values of position and orientation for each of the four nanodiamond/substrate configurations. We find that this is sufficient to clearly capture the behaviour in each case.

## 4.3 Coverslip results and normalisation

Figure 8(a) shows the FDTD calculated (bars) distribution of emission rates for the coverslip geometry relative to the rate $\Gamma_b$ in bulk diamond. The mean calculated spontaneous emission rate is $R_C = 0.12 R_b$, where the emission rate in bulk is $R_b = 1/11.6$ ns [16]. This corresponds to a predicted mean lifetime of $\bar{\tau} = 1/R_c = 96$ ns. The result shows a strong suppression in the emission rate as expected for a dipole enclosed within a dielectric sphere of size much smaller than the emission wavelength [25, 31, 36]. However, both the literature and our own measurements in Figure 6(a) indicate that the experimentally measured mean lifetime of NV centres in nanodiamonds on a coverslip is $\sim \tau_{cov} \approx 20$-$25$ ns [16, 35, 46], which is a factor of 4 shorter than our calculated value.

Given this discrepancy, it is important to validate our calculations. First, as described above, our computational approach reproduces the analytic results for the case of a dipole inside a dielectric sphere and for the case of a dipole near a flat dielectric interface. It involves the same code that has been used successfully in a large number of similar radiation dynamics problems [47-50]. To complement the FDTD calculations, we performed an additional calculation using a semi-analytical approach. Note that in the limit $a \ll \lambda$, the emission rate within an isolated sphere is very weakly dependent on the location of the dipole within the sphere [25]. Therefore we can approximate the sphere-NV centre system as a point dipole with an intrinsic emission rate given by the bulk NV emission rate reduced by the local field factor in Eq. (5). We then use this dipole in the theory of Lukosz and Kunz [33] for emission of a point dipole near a dielectric interface. We calculate the distribution of emission rates over all orientations of the dipole and for the range of positions it can take within the sphere. The results are shown as the red-line histogram in Fig. 8(a). Note that while there is some discrepancy between the two models, which we attribute to the fact that the semi-analytic model ignores coupling between tangential and perpendicular field components which are present in the full model, this semi-analytical approach also predicts relative emission rates much lower than the experimental value of approximately $\tau_{bulk} / \tau_{cov} \approx 0.5$.

We therefore conclude that electromagnetic local density of states effects alone are insufficient to account for the relation between the measured lifetimes of NV centres in bulk diamond and NV in nanodiamonds at a planar surface. This difference may point to some other lifetime-reducing mechanism yet unstudied. Possibilities include a surface damage layer, unknown stress field or strongly non-spherical geometries. We return to this issue in section 5. Nevertheless, our discussion in section 2 shows that the agreement between the measured lifetime and the half-space volume averaging method of Eq. (4), which ignores all polarization effects and the embedding of the dipole in the nanoparticle, must be regarded as coincidental.

Given the considerable disagreement between the mean theoretical and experimental lifetimes, for the remainder of this report we simply concentrate on the relative change in lifetimes for the different substrates, and assume an additional unknown factor gives rise to the absolute lifetime. To this end, the theoretical rates were linearly rescaled so that the calculated mean emission rate for the coverslip corresponds to the experimentally measured value of 25.4 ns. Figure 8(b) shows the distribution of emission lifetimes constructed from the scaled emission rates in Figure 8(a). This scaling does not affect the polarization behaviour which is the focus of this work as we used the same source of diamonds for both the coverslip and opal measurements and the same scaling is expected to apply.

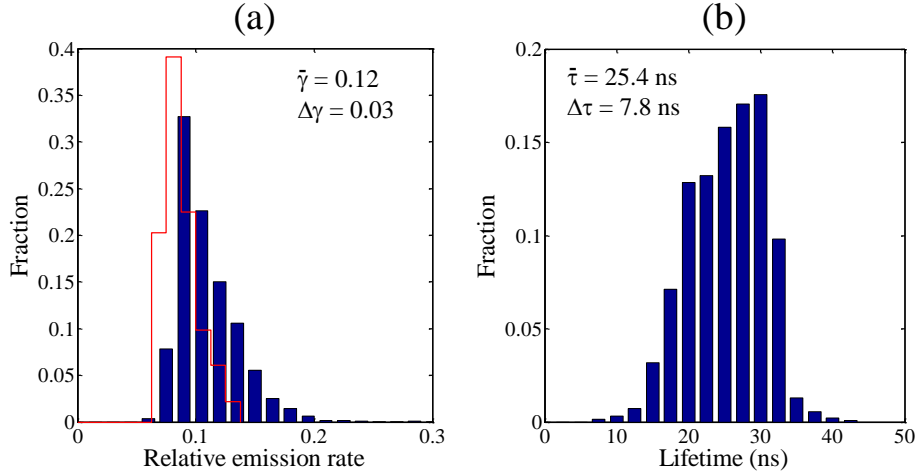

Figure 8: (a) Distribution of calculated FDTD (bars) and semi-analytical (red line) emission rates relative to bulk diamond rate for single NV centres on a coverslip. Quoted values are mean calculated (FDTD) emission rate $\bar{\gamma}$ and the range $\Delta\gamma$ that contains the 50% of centres. (b) Spontaneous emission lifetime distribution after scaling to the measured mean value.

## 4.4 Opal results

Figure 9 shows the calculated lifetime distributions for three indicative rest positions on the opal (interstitial, top and intermediate), as well as a weighted distribution (Fig. 9(d)) based on the probabilities to find crystals in any of the positions as measured by SEM. The weighted case shows a distinct shift to shorter lifetimes ($\bar{\tau} = 15.8$ ns) compared to the coverslip and consistent with the measured reduction of 31% in Figure 6. Note that the distributions for each location are quite different. The mean lifetime for interstitial nanodiamonds (Fig. 9(a)) is a factor of 2 smaller than the mean for nanodiamonds on the top of spheres (Fig. 9(b)). We can understand this significant change qualitatively by reference to the earlier discussion of dipole orientation. In particular, we attribute the reduction in emission rate to the multiple different interfaces of the local environment at the interstitial sites—almost all dipole directions have a component strongly perpendicular to one of the adjacent sphere surfaces. By analogy with the results in Figure 1 for planar interfaces, we might expect that this quasi-perpendicular orientation leads to significant enhancement of the emission. In contrast, for nanodiamonds on the top of spheres, the supporting surface can be regarded as smooth with fairly low curvature and the distinction between parallel and perpendicular orientation remains strong. Note that the difference in emission rates cannot be accounted for merely by a change in the volume fraction $f$ of high index material (polystyrene) in the vicinity of the nanodiamond emitter. If we consider the region within a cubic wavelength of the emitter the volume fraction changes from $f = 0.36$ (sphere top) to $f = 0.46$ (interstitial site). From the calculation of expected lifetimes based only on an index-averaging model in Fig. 2, the reduction in lifetime would only be a factor of 0.04.

The same considerations also help to account for the different relative distribution widths for each location in Fig. 9. For the interstitial and intermediate sites the relative width $w = \Delta\tau/\bar{\tau}$ is $w = 0.57$ and $w = 0.33$ respectively. By contrast, for the sphere top site $w = 0.64$. The dependence of the emission rate on orientation is strongest for planar and near-planar structures, and is minimised for structures where different orientations are somewhat similar as is the case for the interstitial sites. When all possible directions are averaged over, the latter sites must show a weaker variation and therefore a narrower distribution.

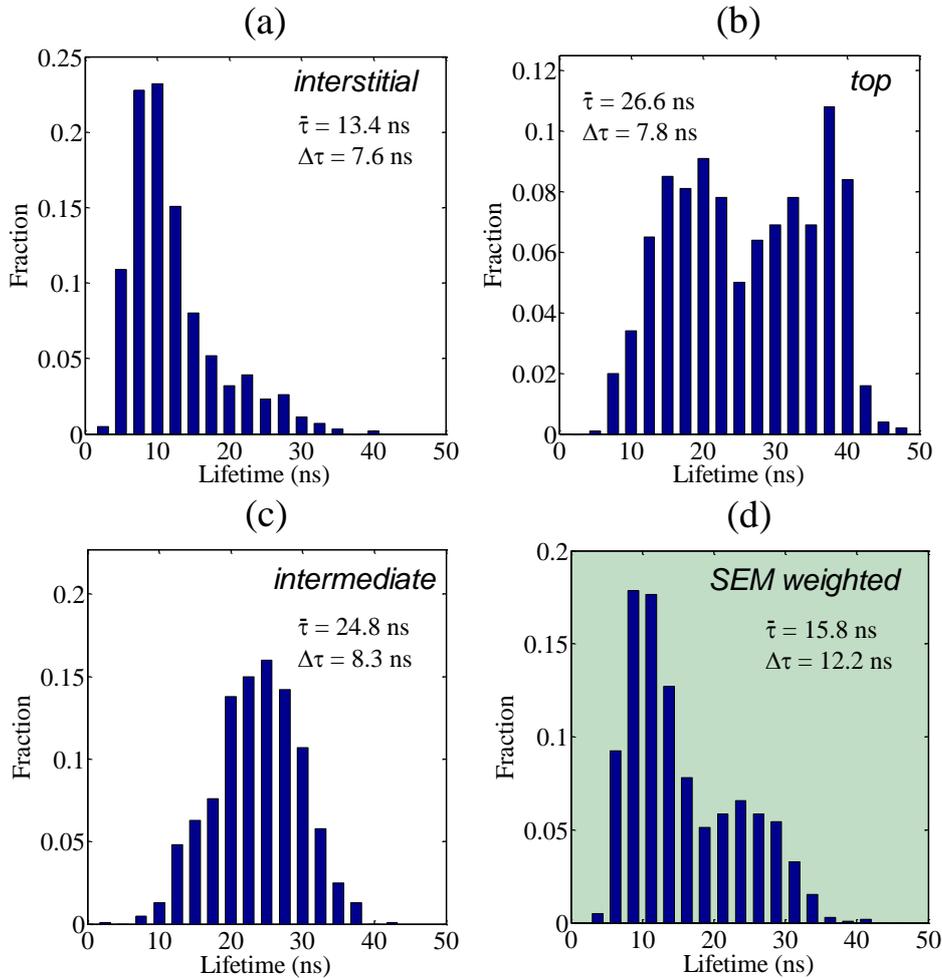

Figure 9: Relative emission lifetime distributions for single NV centres corresponding to various positions on an opal surface along with the overall weighted distribution. Here the weights are provided according to the statistical scanning electron microscopy (SEM) measurements discussed in section 3.1. Each figure shows the mean of the distribution $\bar{\tau}$ and the range $\Delta\tau$ corresponding to 50% of the total centres.

## 5. Discussion

In this section, we consider several additional aspects of our modelling that might be thought to contribute to the discrepancy between predicted and measured lifetimes for diamonds on the coverslip.

Our modelling does not include a variation in crystal size. We performed simulations for a fixed nanodiamond diameter of 54 nm. However, the size variation of the actual nanodiamonds used (discussed in section 3.1) corresponds to the range of $2a/\lambda$ indicated by the green bar in Fig. 7. Over this range, as can be seen, the emission curve for the sphere alone is almost perfectly flat. Thus we expect that the variation in emission rate in the full system due to nanodiamond size should be a very weak effect.

To ensure that our calculated distributions of the emission rate were not peculiar to the choice of spherical nanodiamonds, we performed additional FDTD calculations for a range of nanodiamonds of different shapes supported by the coverslip substrate. The results are shown in Figure 10. We considered a sphere of diameter 60 nm, an octahedron of side length 60 nm resting on a vertex, a cube of side length 60 nm resting on a face, and two oblate ellipsoids with semi-axes lengths $\left(a\eta, a\eta, a/\eta^2\right)$ for $\eta = 1.2$ and 1.4. Despite the range in shapes, the distributions are remarkably similar. The cube shows a significantly broader distribution than the others, because there is a clear distinction between parallel and perpendicular alignment of dipoles that lie close to one of the faces of the cube. For the other shapes, the distinction between parallel and perpendicular orientation is less clear, and the distribution widths are narrower. However, even allowing for the cube case, it is clear that the mean rate and width of the distributions are generic properties of nanodiamonds of this scale and the results in section 4 are not restricted to spherical diamonds alone. While our actual diamonds are not true spheres, Fig. 10 suggests that they would require extreme profiles in order to increase the emission rate by a factor of 4, and this is not supported by SEM analysis.

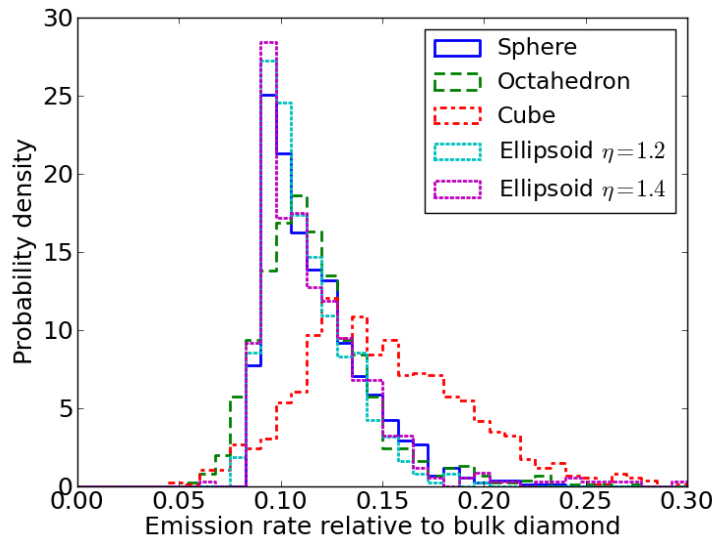

Figure 10: Relative emission rates for nanodiamonds of various idealized shapes on a glass substrate.

Another potential cause of the measured enhancement substantially exceeding the predicted rate might be thought to be enhancement through coupling to guided wave slab modes in the substrate. However, with a coverslip thickness of ~150 μm, it seems very reasonable to treat the substrate as a semi-infinite bulk medium as we have done here.

## 6. Conclusion

Developing a quantitative understanding of the behaviour of single dipole emitters in and around a variety of materials and structures underpins practical single photon device development. We have presented detailed measurements of single photon emission lifetimes for NV centres in nanodiamonds on flat and structured surfaces. The dipole orientation relative to the surfaces around it plays an important role, and the emission rate varies significantly depending on this. We have established that there is a clear difference in lifetime between crystals randomly distributed on a flat surface, and crystals randomly distributed on structured opal surfaces. The experimentally measured single photon emission rates for diamond on the opal show a factor of 1.5 average enhancement compared to the coverslip case. This result is supported by a rigorous FDTD model which indicates that the primary reason for the enhancement is the dipole orientation with respect to the local surface. Comparison of calculations with experiments not only clarify the interaction of an optical dipole in diamond with a surface, but also provides new insights into the physical system which are yet to be understood, including the large discrepancy in measured and predicted absolute lifetimes of NV centres in nanodiamonds. These discrepancies only emerge when the embedding of the dipole in the nanodiamond is explicitly taken into account. In summary, we have presented a comprehensive theoretical model and experimental measurements of enhanced spontaneous emission rates of single NV centres in diamond. Coupled with new advances in fabrication and nano-assembly, this work adds a valuable resource for the development of single photon devices.

## Acknowledgements

A/Prof Rabeau is funded by an Australian Research Council Future Fellowship (FT0991243). F.Ahmed and C. Bradac are supported by MQ Research Excellence International Scholarships and Dr Gaebel is funded by a MQ Research Fellowship. Computer simulations in this work were performed under a merit allocation from the National Computing Infrastructure.